\documentstyle[prb,aps,array,epsf]{revtex}

\begin{document}
\title{The superfluidity mechanism of He II
}
\author{J. X. Zheng-Johansson\footnote{Email: jxzj@mailaps.org}} 
\address{H. H. Wills Physics Laboratory, Bristol University, Tyndall Avenue, Bristol, BS8 1TL, England }
\author{B. Johansson}
\address{Condensed Matter Theory Group, Department of Physics, Uppsala  University, Box 530, 75121 Uppsala, Sweden}
\author{ P-I. Johansson}
\address{Department of Neutron Research, Uppsala University, Studsvik 611 82, Sweden}

\date{April, 2001, submission to Phys. Rev. B}

\def\lsim{{{}\atop{{<\atop\sim}\atop{}}} }
\def\gsim{{{}\atop{{>\atop\sim}\atop{}}} }
\def\lf{\left}
\def\rt{\right}
\def\kbf{{\bf $ k $}}
\def\kbfsu{{\bf $ k \uparrow $}} 
\def\kbfsd{{\bf $ k \downarrow $}} 
\def\oK{$^\circ$K}
\def\e{\varepsilon} 
\def\mef{m_{_{\rm {ef}}}}
\def\G{\Gamma}
\def\g{\gamma}
\def\rbf{{\bf $ r $}}
\def\L{\Lambda}
\def\xbf{{\bf $ x $}}
\def\E{{\cal E}}
\def\Exm{\widetilde{{\cal E}}_x}
\def\Eym{\widetilde{{\cal E}}_y}
\def\Ezm{\widetilde{{\cal E}}_z}
\def\EK{{\cal E}(K)}
\def\Eq{{\cal E}(q)}
\def\Eb{{\bf E}}
\def\Eba{{\bf E}_a}
\def\Ebind{{\bf E}_{ind}}
\def\He4{$^4$He}
\def\Kb{{\bf  K}}
\def\th{\theta}
\def\Tc{ $T_c$}
\def\kb{{\bf  k }}
\def\kbsu{{\bf  k \uparrow }}
\def\kbsd{{\bf  k \downarrow }}
\def\lx{l_x}
\def\ly{l_y}
\def\lz{l_z}
\def\nnr{\nonumber}
\def\Nc{{\cal N}}
\def\ov{\over}
\def\pd{\partial}
\def\ph{{\rm ph}}
\def\a{{\rm a}}
\def\rb{{\bf  r }}
\def\Rb{{\bf  R }}
\def\ub{{\bf  u }}
\def\vb{{\bf  v }}
\def\Tlm{T_{\lambda}}
\def\w{\omega}
\def\wo{\overline{\omega}}
\def\wxm{\widetilde{\omega}_x}
\def\wym{\widetilde{\omega}_y}
\def\wzm{\widetilde{\omega}_z}
\def\xb{{\bf  x }}
\def\xib{\mbox{\boldmath$\xi$}}
\def\d{\partial}
\def\calZ{{\cal Z}}

\def\i{ i}
\def\j{j}
\def\l{ l}
\def\Kcap{ \widetilde{K}}
\def\Kl{K_l}
\def\Kla{K_{l_\alpha}}
\def\Klx{K_{l_x}}
\def\Kly{K_{l_y}}
\def\Klz{K_{l_z}}
\def\Ka{K_{\alpha}}
\def\Kx{K_x}
\def\Ky{K_y}
\def\Kz{K_z}
\def\Kn{K_n}
\def\Kbn{{\bf K}_n}
\def\Kxi{K_{x i}}
\def\Kyj{K_{y j }}
\def\Kzl{K_{z l}}

\def\Kxh{ \widehat{K}_{x}}
\def\Kyh{ \widehat{K}_{y}}
\def\Kzh{ \widehat{K}_{z}}
\def\1N1D{{\cal N}_{1}}
\def\N2{{\cal N}_{2}}
\def\3N{{\cal N}_{3}}
\def\Nth{{\cal N}_{\th}}
\def\uncN{{\cal N}_{uc}}
\def\uN{{\cal N}_{0}}
\def\ueta{\eta_0}

\def\D2{{\cal D}_2}
\def\xD1{{\cal D}_1}
\def\uD{{\cal D}_0}
\def\thD{{\cal D}_{\th}}

\def\Kam{ {\widetilde{K}}_\alpha}
\def\Kxm{ {\widetilde{K}}_x}
\def\Kym{ {\widetilde{K}}_y}
\def\Kzm{ {\widetilde{K}}_z}
\def\kxm{ {\widetilde{k}}_x}
\def\kym{ {\widetilde{k}}_y}
\def\kzm{ {\widetilde{k}}_z}

\maketitle   

\begin{abstract}
{\bf Abstract:} Based on a first principles treatment of the excitation states we show that superfluidity of superfluid $^4$He (He II) results from a reduction in the number of phonon wavevector $K$ states $\N2(K)$ to a level that is negligibly low when the fluid is confined e.g. in a narrow channel, yet wider than the helium atom correlation length, $\Lambda$. This is as a result of the $K$ discretization, a manifestation of the quantum confinement effect (QCE). The predicted relative viscosity of a confined superfluid has the characteristic order of magnitude of experimental data ($<10^{-6}$). Furthermore, we show that at the edges of the resulting energy gaps, the $\N2(K)$ presents discontinuity. When its corresponding energy exceeds the (first) gap, the superfluid flow exhibits a critical velocity $v_c$. Our evaluation of $v_c(d)$ versus the channel width $d$, constrained to satisfy energy conservation, is in good quantitative agreement with experimental data for channels with $d>10^{-6}$ m. Meanwhile, a sharp turn about $K\propto$  $ v_c$ in $\N2(K)$ resembles very well that of the experimental overshoot data. For narrower channels of $d<10^{-6}$ m $\le \Lambda$ in which the phonon excitation picture becomes inadequate, we instead represent the excitation in terms of single atoms with an effective mass, which yields a $v_c(d)$ in close agreement with experiment. Accordingly, the reduction in the number of atomic states results in superfluidity. The theoretical finding in this work, which can be termed the {\bf QCE superfluidity mechanism}, provides a consistent explanation for this puzzling phenomenon, the non-dissipative, superfluidity motion, of He II and could have a significant impact also on the understanding of other superfluids.
\end{abstract}
\noindent
\section{ Introduction}   \label{Secz1}
\noindent
It is one of the most puzzling natural phenomena that in the superfluid phase, He II, and in a narrow channel and below the critical velocity, liquid $^4$He flows as if there is "no friction". Systematic experimental investigations have shown[\onlinecite{WH:Keesom,Allen:Misener:1938,Kapitza,Mehl:etal:1968}] that the viscosity of He II is $10^6 \sim 10^{11}$ times smaller than that of normal fluids. Although the  phenomenon of superfluidity has been the subject of many theoretical studies for the past 70 years, including the well know work by London[\onlinecite{london:1938}], Landau [\onlinecite{landau}] and Feynman[\onlinecite{feynman:1953}], there is still not a satisfactory solution to the problem. In this work, we derive a new theoretical model for the superfluidity of He II based on the relevant experimental indications.
\noindent
\section{  Thermal excitations in the superfluid} \label{Secz2} 
\noindent
To understand the superfluidity, one first of all can learn from the thermal excitation property of the superfluid, for dissipation is caused by thermal excitations. Experiments, in particular inelastic neutron scattering measurements[\onlinecite{pol:1958,henshaw:w:1961,woods:svensson:1978,Svensson:1987,Stirling:el:1990}], have provided basic evidence concerning the nature of the thermal excitations in superfluid He II. Three aspects, in particular, are relevant here. First, the neutron scattering function $S(q,\w)$ of a (pure) superfluid of size ranging from the macroscopic [\onlinecite{woods:svensson:1978,Svensson:1987,Stirling:el:1990}] to the nano scale [\onlinecite{Fak-PRB,Fak-etal,Fak-PRL}] is, at a given momentum transfer $q$ up to at least 1.93 \AA$^{-1}$, sharply peaked at $\hbar \w(q)$, $\hbar \w(q)$ being the neutron energy transfer. For $q \lsim  0.6$ \AA$^{-1}$, the excitations in a macroscopic superfluid are known[\onlinecite{atkins:stasior}] to be well described as  elastic sound waves; here the "macroscopic" or "bulk", which will be assumed for the superfluid till Sec. \ref{Secz5}, refers to a dimension larger compared to the helium atom correlation length. The combined information above indicates that the sound waves in the He II superfluid are quantized; namely in terms of phonons, each having an energy quantum ${\cal E} (K)=  \hbar \w (K)$, where $\hbar$ is the barred Planck constant, and $\w(K)$ and $K$ ($= q)$ are the phonon frequency and scalar wavevector for the isotropic superfluid. Second, $\w(K)$ has been accurately determined from neutron scattering experiments [\onlinecite{henshaw:w:1961}]; in the low $K$ ($\lsim 0.6$ \AA$^{-1}$) region of concern here:
\begin{eqnarray} \label{eq-I1} {\cal E} (K)= \hbar \w (K)= \hbar c_1 K, \end{eqnarray}
$c_1$ being the velocity of first sound of He II. 
Furthermore, thermal conductivity measurement[\onlinecite{Fairbank-Wilks}] has shown that in the superfluid at $\sim$ 1 \oK \ and in the small geometries in question, phonon-phonon collisions are basically absent and the phonons principally collide with the specimen boundaries.  
We can thus effectively represent the propagating phonon wave here by a free "particle" of mass $m_{\ph}$ in real space, for which the de Broglie relation states ${\cal E}= (\hbar K)^2/m_{\ph}$. Combining this with Eq. (\ref{eq-I1}), we get $m_{\ph} = \hbar K/c_1$. The Schr\"odinger equation for the phonon particle, e.g. in the $x$ direction, is $- {\hbar^2 \over  m_{\ph}} {\partial^2 \over \partial x^2}u(x)  = {\cal E}(K_x) u(x);$ its energy is obtained from Eq. (\ref{eq-I1}). The amplitude solution consists of the planewaves 
\begin{eqnarray}\label{eq-I2} u_{x} (K_x) =u_o  e^{i K_x  x}. \end{eqnarray}
 Equations (\ref{eq-I1})-(\ref{eq-I2}), obtained above based directly on experimental observations, give a first principles representation of the excitation states in a macroscopic superfluid sample, and will be the starting point for our subsequent investigation of the confined superfluid. For information only, we mention that the above solutions have been obtained in our microscopic theory of the superfluid of He II[\onlinecite{paperI-II}] through solving the equation of motion of the superfluid atoms (the helium atomic wavepackets). Third, the number of phonon excitations (which can only have occurred between the discrete energy levels of the superfluid) increases in proportion to the concentration of the superfluid in He II and becomes 100\% at zero \oK, \ implying that the superfluid in thermal equilibrium is prominently excited, via phonons. This suggests that insofar as the superfluid motion is concerned, the phonon excitation cannot be arbitrarily omitted and, in arriving at a mechanism of superfluidity, we are obliged to take its effect into full account. 
\noindent
\section{ The theoretical basis of superfluidity motion} \label{Secz3}   \noindent
We consider a superfluid under the condition of thermal equilibrium in steady flow motion along the $z$ axis in a channel of width $d$ and length $L$. Unless specified otherwise, we assume $d<<L$;
this represents a two-dimensional spatial confinement, 2D SC. We also address the cases of one-dimensional and three-dimensional spatial confinements, 1D SC and 3D SC. For simplicity, we assume that $T$ is low ($\lsim$1 \oK) \ so that He II is virtually a pure superfluid.  In this and the next section we discuss wider channels and in Sec. \ref{Secz5} the narrower channels, as the superfluidity mechanism appears to differ in detail for these two cases.

\noindent
\subsection{ Phonon states in a confined superfluid below the first excitation gap} \label{Secz3.1} \noindent
In a superfluid confined to a channel (2D SC) with a $d$ of macroscopic scale, by Sec. \ref{Secz2} the primary thermal excitation consists in phonons. It is readily seen that for this confined fluid (simplified to have a squared cross section) the phonon waves, Eq. (\ref{eq-I2}), are subject to the non-trivial lateral boundary conditions $ e^{i\Kx x}|_{x=0} = e^{i\Kx  x}|_{x=d}$ and similarly for $y$, and along the $z$ axis the trivial one $ e^{i\Kz z}|_{z=0} = e^{i\Kz z}|_{z=L}$, as a result of constructive interference. The phonon wavevector $\Kb$($\Kx,\Ky,\Kz$) in the $x$ and $z$ directions are therefore restricted to the discrete values $\Kbn(\Kxi,\Kyj,\Kzl)$, with 
\begin{eqnarray} \label{eq:apni:7}
\Kxi = {\i \Kxm}, \ \Kyj=\j \Kym,  \quad \i, \j = 
0, \pm 1, \ldots, \pm{(N_{\beta}-1) \over 2}  \end{eqnarray}
$$ {\rm and } \quad \Kzl = {\l \Kzm}, \quad \l = 0, \pm 1,  \ldots, \pm {(N_z-1)\over 2},
\eqno(\ref{eq:apni:7}.b)$$
respectively; $K_n =|\Kbn| = \sqrt{\Kxi^2+ \Kyj^2+ \Kzl^2}$.  Where, 
\begin{eqnarray} \label{eq:apni:8} \Kxm = \Kym 
= {2 \pi \over d}, \quad {\rm and} \end{eqnarray}
$$ \Kzm = {2\pi \over L} \quad\quad\quad\quad\quad  \eqno(\ref{eq:apni:8}.b)  $$
are the smallest $K$ values permitted, the $K$ quanta, in respective direction. $N_{\beta}= {L \over a}$, $\beta=x,y$ and $N_z={d \over a}$, $a$ being the He II interatomic spacing. 
The corresponding minimum excitation energies, or energy gaps, are $ \Delta_{\ph} /\hbar  =  \wxm = \w (\Kx{}_{(\i+1)}) - \w (\Kxi) $ and $ \wzm = \w (\Kz{}_{(\l+1)}) - \w (\Kzl).$
For the 2D SC, the $\Kxi$ and $\Kyj$ states are degenerate. For a 1D SC, the $\Kyj$ and the $\Kzl$ states will be degenerate. For the low energy excitations of Eq. (\ref{eq-I1}), we have 
\begin{eqnarray} 
\label{eq45b} \Delta_{\ph} = \hbar c_1 \Kxm = {h c_1 \over d}, \quad {\rm and}\end{eqnarray}
$$ \wzm = {h c_1 \over L}.   \quad\quad\quad\quad\quad     \eqno(\ref{eq45b}.b)$$ 
Suppose that $d$ is small such that given an external perturbation of energy $\e_s$,
\begin{eqnarray}\label{eq:i:EK}\e_s  < \e_{\ph}^{c}= \sum_{\alpha} \Delta_{\ph},  
\end{eqnarray}
where $ \e_{\ph}^{c}$ refers to the minimal excitation energy per atom permitted by the conditions (\ref{eq:apni:7}); $\sum_{\alpha}$ will be specified in Sec. \ref{Secz3.2}.
Then, no excitations can be caused along the $\Kx$ and $\Ky$ axes. On the other hand, since $L$ is large, $\wzm$ is therefore effectively zero and the low $\Kzl$ modes can undergo continuous excitations.

For the $\w(K)$ of Eq. (\ref{eq-I1}), the number of states in an interval [$\w (\Kn)$, $ \w(\Kn + \delta K) $) is obtained by enumerating the number of $\Kbn'(\Kxi',\Kyj',\Kzl')$ points contained in the spherical shell  $\Kn \le  \Kn' \le  (\Kn+ \delta K)$. Now for the minimum sphere with respect to the $\Kxi$ modes which has the radius
$$K_1 = \Kx{}_{1}= 1 \cdot  \Kxm= \Kz{}_{{L\ov d}}, $$
\input epsf 
 \begin{figure} [as soon as possible]
\begin{center} \leavevmode \hbox{%
\epsfxsize= 7.5 cm  
\epsfbox{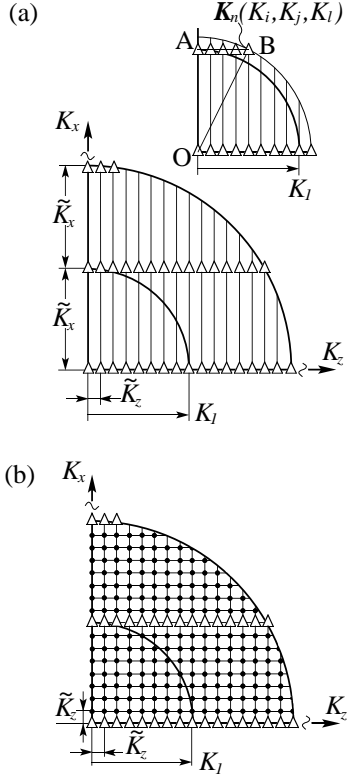}} 
\end{center}
\caption{ 
Configuration of the phonon wavevector $K$ states, shown for $K \le 2K_1=2 \Kxm$ in the first quadrant in the $K_y =0$ plane, for a superfluid conditioned as (a) or (b). (a). The superfluid is 2D confined in a channel of dimensions $d^2\times L$, with $1/(d/L)=8$.  Within $K<\Kxm$, $\Kxm$ being the minimal $\Kxi$ value, the only permitted $K$ states are those along the $\Kzh$ axis indicated by the triangles.  After $K>\Kxm$, an additional number of states are permitted (the triangles) along the $\Kxi=\Kxm$ grid line. The additional states in the first shell $ \Kxm <K \le \Kxm+1 \cdot \Kzm$, indicated by the triangles on the $\overline{AB}$ line in the upper-right graph, give rise to a discontinuous increment in the ${\cal N}_2(K)$ (cf. the inset of Fig. \ref{figzII-2}b).  (b). The superfluid is in a large container of the dimensions $L^3$, being thus un-confined. The permitted $K$ states are indicated by the filled circles in the regions which are unoccupied by the $K$ states of the 2D SC sample of (a), and by the triangles along the grid lines permitted for those of (a). 
 If $1/(d/L)=5 \times 10^4$ as is usually employed in experiments, the number of additional $K$ states -- the filled circles -- of the unconfined sample would be about $10^7$ times more densely populated than illustrated in the figure. \label{fig-Kmesh}  \label{figzII-1}
 }
\end{figure}
\noindent
the only permitted states in it are the $\Kbn(0,0, \Kzl)$ points on the $\Kz$ axis (triangles below $\Kxm$ in Fig. \ref{fig-Kmesh}a). 
Then, accumulating these points in the above manner up to a given $K$ value: $ \Kn = \Kzl= \l \Kzm$ ($\le K_1$), here $n=\l$, gives the total number of phonon wavevector states, or integrated density of states for the 2D SC:
\begin{eqnarray}  \label{eq:appn:6}  
\N2 (K)   = 
 \left\{ \begin{array}{cc}
&\sum_{l'=-l}^{l} \delta_{0 0 l'} = { 2 K \over \Kzm} +1 =2 l +1,  \quad  \mbox{$  K<K_1$ } \cr
&\sum_{l'= - {L/d}}^{L/d}    \delta_{00l'} + 2 \sum_{i'=-1, 1} \delta_{i' 0 0} \quad\quad\quad\quad \  \cr
&\quad\quad\quad\quad\quad  = \lf({2L \over d} +1\rt) +4i|_{i=1},   \quad
                                                                         \mbox{$ K=K_1$}.    \end{array}  \right. \quad\quad\quad\quad  
  \end{eqnarray}
Where, $\delta_{00\l}$ stands for $\delta_{\i \j \l}|_{\i=\j=0}$; $\delta_{\i \j \l}$ equals 1 at each crossing point of the grid lines $\Kxi$, $\Kyj$ and  $\Kzl$, and zero elsewhere. The density of states $\D2(\omega)$ ($={d \N2 (\w(K)) \over d \w (K)} ={\partial \N2 (n) \ov \partial n} {\partial n \ov \partial \w (n)}$) is thus
\begin{eqnarray} \label{eq:dos}
 \D2 (\omega) / (\pd n / \pd \w) =
                         \left\{ \begin{array}{cc} 
                              2,    \quad\quad     & \mbox{$ K<K_1$};         \cr 
                              6,    \quad\quad     & \mbox{$ K=K_1$}.          \end{array}  \right.   \end{eqnarray} 
where $\partial n /\partial \w  = L / 2 \pi c_1 = 1/\Kzm c_1$. The $\N2(K)$ states, which can be continuously excited, will strictly speaking dissipate the flow motion in question, however, only to an extremely small extent.  One way of seeing the small extent is by comparing with a large (unconfined) sample of the dimensions $L^3$.  For the latter, the allowed phonon states are similarly defined by boundary conditions as Eqs. (\ref{eq:apni:7}) except that $d$ is replaced by $L$. Accordingly the $K$ states for it are defined by the crossing points (filled circles in Fig. \ref{fig-Kmesh}b) of the grid lines $(\Kxi,\Kyj,\Kzl)$, equally spaced by a $\Kzm$ in all axial directions, the volume quantum being ${\Kzm}^3$. Within a sphere of radius $K = n \Kzm$, of a volume $\Omega = {4 \pi \over 3} K^3$, the total number of states and density of states for the unconfined sample are
\begin{eqnarray} \label{eq:unc:sta}
\uN (K) = {\Omega \over {\Kzm}^3} = {4\pi \over 3} n^3, \quad {\rm with } \ \uN (K_1) = {4\pi \over 3} \left({L \over d}\right)^3;  \\
\label{eq:unc:dos} 
{\rm and} \quad  \uD(\omega) /(\pd n / \pd \omega) = 4 \pi n^2. \quad\quad\quad\quad\quad\quad\quad\quad\quad\quad \end{eqnarray}
Comparing $\N2(K_1)$ of Eq. (\ref{eq:appn:6}) with the $\uN(K_1)$ above,
we obtain that up to $K=K_1$ the number of states for the 2D confined sample is reduced by a relative amount (noting $L/d >>1$, $2$):
\begin{eqnarray} \label{eq:i:5-2} r_{2,0}= { \N2(K_1) \over \uN (K_1)  } 
\simeq {3 \over 2\pi }  {1 \over \left( {L \over d }\right)^2 }.  
\end{eqnarray}
As can be seen, $r_{2,0}$ (solid line in Fig. \ref{fig-dos}a)  is very small for $L/d >>1$.
 For $L/d \sim 5 \times 10^4$, which is a typical sample dimension ratio used in superfluidity experiments[\onlinecite{Allen:Misener:1938,Kapitza,Mehl:etal:1968}],  the number of phonon states is reduced by a factor $r_{2.0} \sim 10^{-10}$, to $10^{10}$ times smaller. As we also see, the extremely small $\N2(K)$ corresponds to the removal of those states (filled circles in Fig. \ref{fig-Kmesh}b) which are permitted for the unconfined sample, lying in the interim spanned by the discrete states $\Kxi, \Kyj$ of the 2D SC sample. It is noteworthy that the degeneracy of each discretized state is not increased.  
The extraordinary reduction in $\N2(K)$ in the reciprocal $K$ space is also as expected directly from its relationship with the real space. We can readily inspect this relation by multiplying the maximum numbers of modes in all three axial directions given in Eqs. (\ref{eq:apni:7}); 
we get the total number of states: $\Nc_{\th}(K_{max}) = N_x N_y N_z ={V_{\th} \over a^3}=N_{s \th}$, $N_{s \th}$ being the number of atoms in a volume $V_{\th}$ under $\th$D SC. This equality says that, reducing $V_0=L^3$ to $V_{2}=d^2L$ corresponds to removing the atom population by the factor ${N_{s2}\over N_{s0}}= {V_0\over V_2}=\left({d\over L}\right)^2={\N2 \over \uN}$ -- the same conclusion as Eq. (\ref{eq:i:5-2}); the simple evaluation above shows in addition that $\N2(K)$ is reduced over the entire $K$ volume.

\noindent
\subsection{  The superfluidity mechanism} \label{Secz3.2}  \noindent
Consider a superfluid bulk of $N_s$ atoms and total mass $M_s=m N_{s}$, $m$ being the $^4$He atomic mass, set to flow motion at velocity $v_s$. Its translation momentum and energy are thus $P_s = M_s v_s$ and $  E_s =  {1\over 2}M_s v_s^2$. Suppose now there presents a flow-wall interaction, or a viscous force. The flow energy will then be gradually dissipated and converted to an additional, disordered thermal vibration energy of the atoms, which by Sec. \ref{Secz2} are in the form of phonons, in the interfacial layers.  The total flow-wall interaction consists of the collision of all of the interfacial atoms from the superfluid side, denoting their total number by $N_A$, with all of those on the wall side. The flow energy per collision atom, as the $\e_s$ in Eq. (\ref{eq:i:EK}), is thus,  
\begin{eqnarray}\label{eq-Es}
  \e_s =  { E_s\over N_A}= {1\over 2}\mef v_s^2;  
\end{eqnarray} 
 This physically represents that the superfluid atoms, being strongly correlated, will collide with the wall as if each has an effective mass: 
\begin{eqnarray}\label{eq45ef} \mef=\left({N_s \over N_A}\right) m. \end{eqnarray} 
\input epsf
  \begin{figure}[top]
\begin{center} \leavevmode \hbox{%
\epsfxsize= 7.9 cm  
\epsfbox{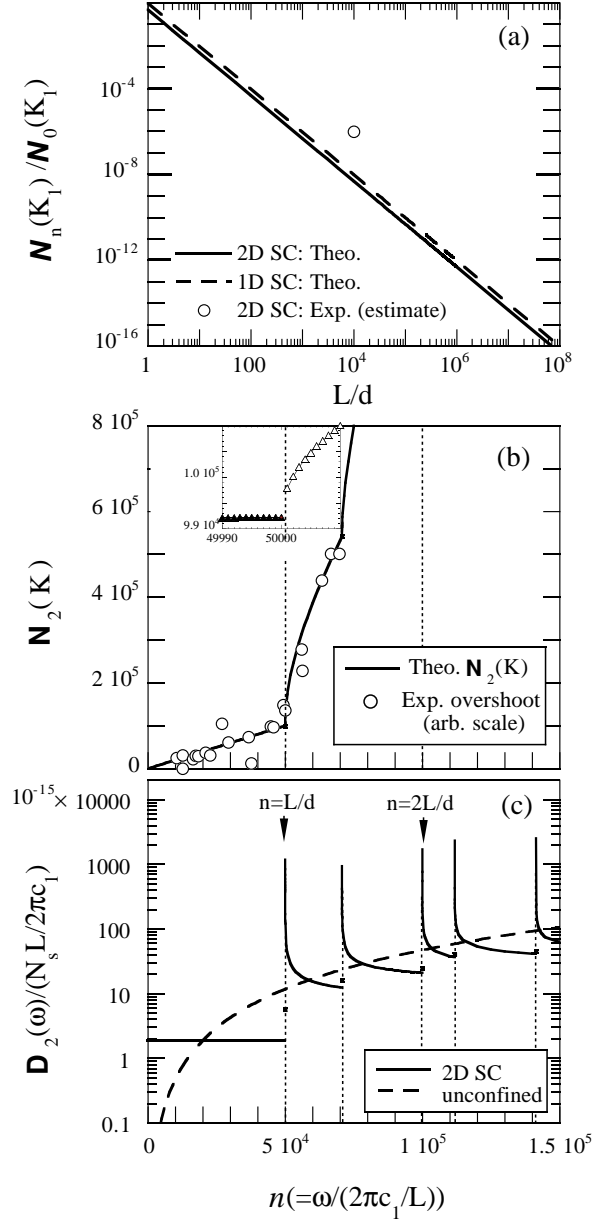}} 
\end{center}
\caption{ 
(a). Ratio of the integrated phonon destiny of states of a $\th$D confined superfluid, $\Nc_{\th}(K_1)$, over that of an unconfined superfluid $\Nc_{0}(K_1)$, as plotted as a function of $L/d$; $d$ and $L$ are the smallest and largest dimensions of the confined sample and $L^3$ of those of the unconfined.  
$\th=1$ or 2 refers to a 1D or 2D SC. (b). The theoretical result for $\N2(K)$ v.s. $l $ plotted up to a $K= 2\Kxm$ and hence $\l =2\Kxm /\Kzm = 2L/d$, with $d/L= 5 \times 10^4$; $\l =L/d$ and $2L/d$ are the first and second lowest $\Kxi$ states. $\N2(K)$ has discontinuous increments at $i\Kxm, i=1, 2, \ldots$, the discontinuity at $K_{1}$ is shown on an enlarged scale in the inset. The circles indicate the experimental overshoot data (from Ref \ref{Winkel:etal:1955}), $O_{\exp}$, versus flow velocity $v_{s.\exp}$. $O_{\exp}$ is scaled to $O_{{\rm plot}} = {9.9 \times 10^4 \over 8} O_{\exp}$ in the graph, so as to fall within the magnitude of $\N2(K)$. Meanwhile, $v_{s.\exp}$ is scaled to ${5 \times 10^4 \over 15} v_{s.\exp}=\l$ so that the turning point of the experimental curve (i.e. $v_s \simeq 15 =$ cm/sec $v_c^{\exp}$) occurs at $\l = L/d= 5 \times 10^4$.  Despite of the arbitrary scaling, the valid information is that the variation of the theoretical $\N2(\l)$ v.s. $\l$ at about $\l = L/d$ closely resembles that of the experimental data: both $O_{\exp}$ and  $\N2(K)$ exhibit a smooth increase before $\l$ reaching the critical velocity (corresponding to $\l= L/d$) and after that a much more rapid increase. 
(c). The corresponding phonon density of states $\thD(\omega)$, divided by a constant $N_s L/2\pi c_1$, for the 2D confined superfluid ($\th=2$, solid curves and circles) and an unconfined superfluid ($\th=0$, dashed curve). The dotted thin lines are guide for the eyes.  \label{fig-dos}  \label{figzII-2}
} \end{figure}
\noindent
On the other hand, since the inert character of the helium atoms upon entering the He II state should not change with respect to the wall, we can regard that the flow atoms (with an effective mass), on passing a wall atom, do not interact with it except at the moment of collision. Then, the maximum excitation energy, $\w(K(v_s))$, will be directly proportional to $v_s^2$ for the flow-wall inelastic collision here; see further Sec. \ref{Secz4.2}. A successful excitation is dictated by the following factors. 1). The number of available states for excitation at each accessible $K$ value, which is represented by $ \thD (\w(K))$. When its $\e_s$ satisfies (\ref{eq:i:EK}), the flow momentum, being along the $z$ direction, will directly convert to the only allowed $K_z$ phonon momentum. On the other hand, when $\e_s$ is able to excite the  $\Kzl = i\Kxm$ modes ($i \ne 0$), the excess phonons, produced first in the $z$ direction, will cause a momentary stress difference between the $z$ and the lateral ($x$, $y$) directions, and hence a momentum flow from the former to the latter. This will continue until, presumably within a relaxation period, the excess momentum is uniformly distributed amongst all the degenerate states (within a quantum uncertainty range), as dictated by the isotropy of the fluid. 2). The probability of excitation of a phonon of $\w(K)$ on perturbation. This is determined by the Boltzmann factor $f(\w, T) =\exp(-\hbar \w/k_B T)$, assuming a one-phonon process for the low temperature involved. The sum of $f(\w, T)$ over all excitation events in a certain time interval $\Delta t$ should equal $[<n>(T+\Delta T)-<n>(T)]$, $<n>(T)$ being the Planck distribution at $T$, and $\Delta T$ the temperature increment due to the excess phonons in $\Delta t$. 3). The rate of head-on collisions in which a flow atom loses its entire translation energy. We define $\g$ as the number of head-on collisions an interfacial atom commits per unit travelling-distance, and hence $\G=\g \ell$ that after travelling a distance $\ell$. 4). There present in the interfacial layer $N_A$ simultaneous collisions at any one time. Multiplying the above factors and summing over all accessible states, we then obtain the total excess excitation energy produced by the flow-wall interaction over a travelling distance $\ell$ in a channel of $\th$D SC: 
\begin{eqnarray} \label{eq-Nph}
& E_{\ph.\th}(\w(K)) &  = \G N_{A.\th}\sum_{\w'\equiv\w(K') \le w(K)} \hbar \w'  \thD(\w')  f(\w',T) \nonumber \\
&      &  \simeq \G N_{A.\th}  \hbar  \wo (K) f(\wo(K),T)  \sum_{\w' \le w(K)} \thD(\w')  \nonumber \\
&        & = \G N_{A.\th}  \hbar \wo (K) f(\wo (K),T) {\cal N}_{\th}(K).  \end{eqnarray}
 In going to the second expression on the right hand side the $\w' f(\w',T)$ is for simplification replaced by its intermediate value at $\wo \approx {1\ov 2} (0+\w)$, on the grounds that in the narrow range (0, $w(K_1)$), $\w f(\w,T)$ varies insignificantly. The $E_{\ph.\th}(\w)$ above is equal to the loss of $E_s$ over the distance $\ell$: $E_s(\ell)-E_s(0)$. Now, the average viscous force, $F_\th$, is determined by $F_\th= E_{\ph.\th} / \ell$, where the $E_{\ph.\th}$ as defined in Eq. (\ref{eq-Nph}) in turn equals the total dissipative work ${E_s(\ell)-E_s(0) \over \ell}$ done by the wall to the flow. Next, by its definition, and by combining with Eq. (\ref{eq-Nph}), the viscosity between the wall and the flow is then obtained as
\begin{eqnarray} \label{eq-eta}
& \eta_\th &= {F_\th \over A_\th \cdot ({\delta v \ov \delta x})}  ={ E_{\ph.\th} \over \ell A_\th \cdot  ({\delta v_s \ov \delta x})} \nonumber \\
 &           &= {\G \hbar \w(v_s) f(\wo (v_s))  {\cal N}_\th (\w) \over \ell a^2 \cdot ({\delta v \ov \delta x})}
\end{eqnarray}
where ${\delta v \ov \delta x}$ ($\sim {v_s-0 \over a}$) is the lateral velocity gradient and $A_\th =N_{A.\th} \cdot a^2$ the area of the flow-wall interfacial layer. Furthermore, superfluidity experiments have commonly shown that an unconfined superfluid exhibits essentially a normal viscosity which may be represented here by that of He I, denoted by  $\eta_{{\rm He I}}$. Thus, $\ueta \simeq \eta_{{\rm He I}}$. Comparing the equations (\ref{eq-eta}) for $\th=2$ and $0$ respectively for a fixed $\ell$ and placing in them the preceding relations, we then obtain the relative viscosity of a 2D SC superfluid flow, namely the ratio of $\eta_{2}$ to $\eta_{_{{\rm He I}}}$, to be 
\begin{eqnarray} \label{eq:eta:1}
  { \eta_{2} \over \eta_{_{{\rm He I}}}}  \simeq  { \eta_{2} \over \ueta } 
= {\N2(K) \over \uN(K)  } ={3 \over 2 \pi} \left({1 \over {L / d}}\right)^2. \end{eqnarray}
 In going from the second to the third expression on the right hand side above, the parameters $\G$, $\ell$, ${\delta v \ov \delta x}$, $a^2$, and $f(\wo, T)$ common to the samples of $\th=2$ and $\th=0$ cancel out, yielding a $r_{2,0}$ as of Eq. (\ref{eq:i:5-2}). This is to say that, the extraordinary reduction of the number of states of a confined superfluid as observed in Sec \ref{Secz3.1} in effect accounts for the reduction in its viscosity.  For $L/d \approx 10^3 \sim 5 \cdot 10^{4}$:
$${ \eta_{2} \over \eta_{_{{\rm He I}}}} \lsim 10^{-7} \sim 10^{-10},  \eqno(\ref{eq:eta:1})'$$
which represents for the given dimension of the confined He II sample the lowest relative viscosity, namely that due to the pure superfluid. In the superfluidity experiments where the superfluid sample used has a dimension ratio $L/d \sim 5 \cdot 10^{4}$, it has been estimated[\onlinecite{Allen:Misener:1938,Kapitza,wilks}] that the viscosity of the superfluid is lower by at least a factor $10^{-6}$ compared to that of a normal fluid, or, that of He I; i.e. $\eta_{2}^{\exp}/ \eta_{{\rm He I}}^{\exp}\simeq 10^{-6}$ (circle in Fig. \ref{fig-dos}a).  In comparison, our theoretical value (solid or dashed line in Fig. \ref{fig-dos}a) is well below this.  Since a small portion of the normal fluid component would contribute to the experimentally measured viscosity, a somewhat lower theoretical value for the pure superfluid as obtained above is as expected. Therefore, we may conclude that the theoretical $\eta_{2}$ reproduces the characteristic order of magnitude of experimental data.

For a 1D confined superfluid, the relations corresponding to Eqs. (\ref{eq:appn:6}) and (\ref{eq:i:5-2}) at $K=K_1$ are:
$ \1N1D(K_1) = {4 L\over d} +3, $ and
$ r_{1,0}  = {3 \over \pi} \lf({1 \over L/d}\rt)^2 $
respectively. This says that, similar to $\N2(K_1)$, the reduction of $\1N1D(K_1)$ is, too,  proportional to $(d/L)^2$; and $r_{1,0}$ (dashed line in Fig. \ref{fig-dos}a) is only twice larger than $r_{2,0}$, which makes no essential difference to the absolute scale of $r_{2,0}$. In the case of 3D SC, $\3N(K<K_1) \equiv 0$ and hence $\eta_{3}\equiv 0$. Any measurable viscosity should therefore result from the normal fluid component of He II. The 3D SC may be compared to the superfluidity experiments where He II flows through porous materials[\onlinecite{Mehl:etal:1968,Alphen:etal}]; although experimental viscosity of He II (below $v_c$) so far given is not well separated as to resulting from which component. It is suggestive however to observe that in the experiment by Mehl et al.[\onlinecite{Mehl:etal:1968}] who employed a 3D SC, the ratio $\eta_{2}^{\exp}/\eta_{{\rm He I}}^{\exp} $ was estimated[\onlinecite{wilks}] to be at least $10^{-11}$, which is an order of $10^{-5}$ better than the estimated ratio $\eta_{2}/\eta_{{\rm He I}} <10^{-6}$ for the capillary experiments [\onlinecite{Allen:Misener:1938,Kapitza}] with a similar $d$ value.

The theoretical evaluation of viscosity above, supported by its good agreement with experimental results, has already underlined a superfluidity mechanism of $^4$He which we now summarize as follows:  
{\it In a superfluid subject to a 2D or 1D spatial confinement with a dimension ratio $d/L$, the number of phonon states $\Nth(K)$ ($\th=2$ or 1) will be reduced to the fraction $(d/L)^2$ of that of a unconfined sample, and is effectively zero for $d/L<<1$. Whereas for a 3D SC, $\3N (K) = 0$. In any of the cases, the flow motion of the superfluid with $\e_s$ satisfying Eq. (\ref{eq:i:EK}) will be effectively "non-dissipative" on the time scale during which a normal fluid flow velocity will have long been exhausted. This mechanism has its origin in Quantum Confinement Effect (QCE), and will be referred to as the {\bf QCE superfluidity mechanism}. 
}

That the QCE is responsible for the superfluidity motion of He II is further directly pointed to, in the first instance, by the experimentation by Reppy et al.[\onlinecite{JD:Reppy}] and others[\onlinecite{Mehl:etal:1968,Henkel:etal:Reppy:1968}], which showed that after being initially set to rotation in a vessel, a He II current persisted typically for hours. This clearly reveals that the "persistency" has to be of a "non-dissipative" origin, or, the dissipation is negligible at least within a duration of hours.  In the second case, a wide range of experiments show that He II actually displays normal friction in a macroscopic bulk flow, its non-dissipative superfluidity only presents in a narrow channel -- direct evidence of the confinement effect -- and below a critical velocity, $v_c(d)$ which is inversely proportional to $d$ (circles in Fig. \ref{fig-vc}). The good quantitative agreement between the theoretical value of $v_c(d)$ to be obtained in Sec. \ref{Secz4.2} and experimental data will add one further support to the QCE superfluidity mechanism. Indeed, it is also as expected from basic principles of quantum mechanics, that QCE is deemed to play a role in the superfluidity phenomenon, which does occur in  liquid helium when it turns into a quantum liquid, the He II, and is spatially confined.
\noindent
\section{ Critical velocity $v_c$.} \label{Secz4} 
\subsection{ \label{Secz4.1}   Prediction of the presence of a $v_c$}  \noindent
When the $\e_s$ of Eq. (\ref{eq:i:EK}) begins to just exceed $\sum_\alpha \Delta_{\ph}$,
the states satisfying $K> \Kxm$ will be progressively accessible to excitations.  Make reference to the geometric relation indicated by the large triangle $\overline{OAB}$ (upper graph of Fig. \ref{fig-Kmesh}a), where $\overline{AB}_{i=1} =\sqrt{\lf(n\Kzm\rt)^2 - \lf(\i \Kxm\rt)^2}$ and $n={K\ov \Kzm}$, with $\i=1, \j=0$  and $\nu=n - \i  {L\ov d}=1$ in the plot. Similarly, a grid line of ($i,j$) $\ge$ ($1,1$) (not shown in the figure) has the length $\overline{CD}=\sqrt{\lf(n\Kzm\rt)^2 - \lf(\sqrt{\i^2+\j^2} \Kxm\rt)^2}$. It is straightforward that the $\N2(K)$ and $\D2(\w)$ within $K \le \Kn =\i \Kxm + \nu \Kzm = n \Kzm$ (where $\nu=1, 2, \ldots, {L\ov d}$) can be obtained to be
\begin{eqnarray}
 &  \N2(K)  &=  2 \sum_{\l'= - n}^{n}    \delta_{00 \l'}  +
\sum_{{(\i',\j',\l') \atop =(0,0,1)}}^{(\i,\j,\l)} \delta {}_{\i' \j' \l'} =  [\N2(K_1) + 2\nu]  \quad\quad\quad\quad \nonumber \\
& &   + { \lf[4i+8\sum \overline{AB}_{i'}i\rt] + 
 \lf[\sum g_{i',j'}(4+8 \overline{CD}_{i,j})\rt] \ov \Kzm}  (\ref{eq:appn:6}.b) \quad\quad\quad\quad \nonumber \\
 & &  = [2 n + 1] + \lf[  4 \i  |_{n=i'{L\ov d}} + 8 \sum_{\i'=1, n\ge i'{L\ov d}}^{i'=i}  \sqrt{n^2- \left(\i' {L\over d}\right)^2}+  
 \rt]  \quad\quad\quad\quad \nonumber \\
 & &  +  \sum_{i'=1, i'\ge j', \atop  n \ge \sqrt{i'^2+j'^2}{L\ov d}}^{i'=i} 
 g_{i',j'}\lf[ 4+ 8 \sqrt{n^2-\lf(\sqrt{i'^2+j'^2}\ {L\ov d}\rt)^2 } \rt], 
\nonumber \qquad\qquad \qquad \qquad \qquad \qquad \qquad (\ref{eq:appn:6}.b) 
 \end{eqnarray}
\noindent 
and  
\begin{eqnarray}
 \D2\lf(\omega(K)\rt) /({\d n \ov \d \omega}) =2 + 4|_{n=i}+ \sum_{{\i'=1, \atop n>i'{L\ov d}}}^{i'=i} {8 n \over \sqrt{n^2 -
\lf(\i'{L\over d}\rt)^2}}  \quad\quad\quad \nonumber \\ 
+ \sum_{ i'=1, i'\ge j', \atop n>\sqrt{i'^2+j'^2} {L\ov d}}^{i'=i} g_{i',j'} {8n \ov \sqrt{n^2-\lf(\sqrt{i'^2+j'^2}\ {L\ov d}\rt)^2}} + 4g_{i,j}|_{{n=\atop \sqrt{i^2+j^2}}},  \quad\quad\quad \nonumber 
\end{eqnarray} 
$$
   \eqno{{\rm where} \  \i K_1<K < (\i+1) K_1, \quad\quad (\ref{eq:dos}.b)}$$
and $g_{i,j}={1\ov 1}, {2\ov 1},{2\ov 2}, 2,2,3$, for $(i,j)=$  $(1,1)$, $(2,1)$, $(2,2)$, $(3,1)$, $(3,2)$, $(3,3)$.   For $\i=1$, the above simplify to:
$$ \N2(K) = (2n + 1) + 4\cdot i|_{i=1} + 8 \sqrt{n^2- \left( {L\over d}\right)^2} \quad \eqno{(\ref{eq:appn:6}.b)'}  $$
\begin{eqnarray} \quad    + \lf\{  \begin{array}{c}  0, \quad\quad\quad\quad K_1 <K \le \sqrt{2}K_1 \\    
       4\cdot i|_{i=1}  +  8\sqrt{n^2-(\sqrt{2}{L\ov d})^2}, \quad   \sqrt{2}K_1<K\le 2K_1, \quad \quad\quad\quad\quad\quad\quad\quad\quad\quad
             \end{array}  \rt. \nonumber 
\end{eqnarray}
and 
$$  \D2(\omega) /(\d n / \d \omega) =6 + {8 n \over \sqrt{n^2 -
({L\over d})^2}}     $$
\begin{eqnarray} \quad    + \lf\{  \begin{array}{c}  0, \quad\quad\quad\quad\quad\quad   K_1 <K \le \sqrt{2}K_1 \\    
         4 + {8n \over \sqrt{n^2-(\sqrt{2}{L\ov d})^2}}, \quad   \sqrt{2}K_1<K\le 2K_1.
             \end{array}  \rt. \nonumber  \qquad\qquad\qquad\qquad\qquad\qquad         (\ref{eq:dos}.b)'
\end{eqnarray}
For ${1 /(d/ L)} = 5 \times 10^4$, the resulting $\N2(K)$ and $\D2(\omega) {1\over (\d n / \d \w) N_s} $ are graphically shown in Fig. \ref{fig-dos}b-\ref{fig-dos}c, where $N_s$ $= {d^2L\ov a^3}= 1.07 \times 10^{15}$.
$\N2(K)$ is a stepwise function (Fig. \ref{fig-dos}.b); it first increases discontinuously at $K=\Kxm$ by a small amount $\delta \N2(\Kxm) = 4 $, and then at $K=\Kxm+1 \cdot \Kzm$ by a large amount: 
\begin{eqnarray} \label{eqdN} &\delta \N2 (K_1+\Kzm) &=  \N2(\Kxm+\Kzm) - \N2(\Kxm) \nonumber 
\\ &&=\sum_{\l'=0}^{L/d+1} {\delta_{1 1 \l'}} = 8 \sqrt{ {2L\over d} +1 }.
\end{eqnarray}
i.e. $\delta \N2(K_1)$ amplifies in proportion to $\sqrt{L/d}$; $\Kzm$, being $<<\Kxm$, is dropped from $\delta \N2$. For the given $ L/d$, $\delta \N2(K_1)$ is as large as $\sim 2.5 \times 10^{3}$ per atom (inset of Fig. \ref{fig-dos}b). After this stepwise turning, $\N2(K >K_{1}+\Kzm)$ rises much more rapidly with $K$ as compared to $\N2(K<K_{1})$. Similarly at each higher $\Kxi =\i  ({L\ov d}) \Kzm$  (${\i=1,2,\ldots, \atop {\rm and}\  j=0}$) or $K_n =\sqrt{i^2+j^2}  ({L\ov d}) \Kzm$ ($(i,j)=(1,1), \ldots$)value, $\N2(K)$ has a discrete jump by $\delta \N2 (\i \Kxm)$.  These above discontinuous jumps of $\N2(K=i\Kxm)$ correspond to sharp peaks in the $\D2(\omega(K)) $ v.s. $n$ function (Fig. \ref{fig-dos}c).  Similarly, for a 1D SC at $K=n \Kzm$ and $\i=1$:
$  \1N1D(K) = \sum_{{\j',\l' = \atop -n,-n}}^{n,n} \delta_{0 \j'\l'} +
 \sum_{ \j',\l'=0, 1 \atop \i'=1 }^{\j',\l'=n, n} \delta_{\i' \j' \l'}
=4 n + 3 +  4 \sqrt{n^2- \left({L\over d}\right)^2}$. 
$\xD1$ can be then similarly obtained. It can be noted that $\N2(K>K_1)$ or $\1N1D(K>K_1)$ is still extremely small when compared to $\uN(K>K_1)$.   

From Eq. (\ref{eq-eta}), it directly follows that $\eta_{2}$ will behave analogously as $\N2(K)$ described above. Such behaviour indeed resembles that of thermal excitation related quantities as observed experimentally, including pressure difference[\onlinecite{Daunt:Mendelssohn:1946,Atkins:1950,Bowers:Mendelsson:1950}],  or temperature difference[\onlinecite{Brewer:Edwards:1961}] across a flow that is required to produce a small increase of the flow velocity, or both [\onlinecite{Hung:etal:1952,Winkel:etal:1955}], which have shown to rise rapidly when $v_s$ begins to exceed $v_c$. The close resemblance is illustrated by plotting in Fig. \ref{fig-dos}b
the experimental overshoot data (the amount of He II flux driven by the temperature difference across the channel generated by a previous heat supply) from Ref \ref{Winkel:etal:1955} (circles) together with the theoretical $\N2(K)$ indicated by the solid curve. If suppose that now a constant pressure difference is maintained to increase the flow velocity. Then, by (\ref{eq:appn:6}.b), when $v_s $ exceeds a $v_c$ corresponding to $\Kxm$ the flow will receive a sudden deceleration so as to appear as if suddenly disturbed. This actually resembles another typical phenomenon observed in experiments[\onlinecite{wilks}].
As we indicated in Ref. [\onlinecite{gap-reslo}], a direct measurement of the peaked structure in $\D2(K)$ (in the low $K$ region) using neutron scattering is not presently feasible at least for a macroscopically confined He II, as is limited by instrument resolution ($\delta q$ $\gsim$ 0.005 \AA$^{-1}$ and $\delta \w \gsim$ 0.01 $\mu$eV being roughly the highest achievable today) compared to the small $\Kxm$ and $\Delta_{\ph}$ involved; and similarly also for a microscopically confined He II if lower resolution is used. In the case of the recent neutron scattering experiments[\onlinecite{Fak-PRL}], where He II is e.g. confined in pores of size $\sim$70 \AA \ in Vycor, the resultant $\Kxm$, $\sim$0.09 \AA$^{-1}$, cannot be resolved with the $q$ resolution $\gsim$0.4 \AA$^{-1}$ used in these experiments. Furthermore, although with a high resolution instrument the $\Kxm$ and $\Delta_{\ph}$ due to pores $\lsim$ 100 \AA \ can in principle be resolved, however the possible effect from the varying size and shape of the pores in the porous material, which if non-trivial will smear the information from each single discrete state, would need be first investigated.

\noindent
\subsection{  Exact evaluation of the $v_c(d)$ }  \label{Secz4.2}  \noindent
Consider now the flow velocity is increased such that the $\e_s$ of each collision atom is just able to produce an excitation of the states at $K_1$; by Sec. \ref{Secz4.1}, here $v_s$ has its critical value. For the critical condition in question here, we only need to consider head-on collisions and, of these, the successful excitations. Then, the threshold excitation energy is given by the summation of Eq. (\ref{eq-Nph}) over only the degenerate modes at  $K=K_1$:
\begin{eqnarray}  \label{eq-Ecr}
& \e_{\ph.2}^{c}  &= {(E_{\ph.2} (K_1)-E_{\ph.2} (K_1-\Kzm)) \over \G N_{A.2}   f(\wo(K),T)} \nonumber \\ 
&             &=   \sum_{\w' = \w(K_1)} \hbar \w' \D2(\w')
\end{eqnarray}
where $ \sum_{\w' = \w(K_1)} \hbar \w' \thD(\w')=\sum_{\alpha} \Delta_{\ph}=\e_{\ph}^{c}$ is as in Eq. (\ref{eq:i:EK}); $\alpha=\pm(x,y,z)$.
On the other hand, for a flow atom in collision to be able to produce the minimal excitation above, in the severest case the atom has to lose its entire translation energy as given by Eq. (\ref{eq-Es}) and be decelerated from $v_s=v_c$ to $v_s=0$. Combining Eqs. (\ref{eq-Es}) and (\ref{eq-Ecr}), we therefore obtain the threshold relation for the flow to thermal energy conversion, or the condition for the onset of critical velocity $v_c$ ($=(v_s)_{min}$),  expressed for per collision atom to be
\begin{eqnarray} \label{eq3-7b}
0- {1\over 2}\mef v_c^2   = -\sum_{\alpha} \Delta_{\ph}, \quad {\rm or} \quad  \e_s^{c} = \e_{\ph}^{c}. \end{eqnarray}
Two remarks may be made here. First, the collision atom having a large effective mass represents a classical particle; its collision with the wall, being inelastic, is thus governed only by the energy conservation condition in Eq. (\ref{eq3-7b}). This is in contrast to the un-correlated quantum helium atoms in Sec. \ref{Secz5}. Second, the "stopped" atom as in Eq. (\ref{eq3-7b}) will exchange momentum with its surrounding atoms in the flow thereby reducing the total flow translation energy; since on satisfying Eq. (\ref{eq3-7b}) many atoms at a time will be "stopped", this will then decelerate the flow motion abruptly. 

Suppose now the channel of a cylindrical shape has a diameter $d$ and length $L$, and thus an interfacial area of $A= \pi d L$ and volume $V= L {\pi d^2\ov 4}$. 
We thus have  $N_{A}={A\ov a^2} = {\pi d L\ov a^2}$, 
$M_s = N_s m = {V\ov a^3}m= {\pi d^2 L \ov 4 a^3} m$, and 
$$\mef = {M_s \over N_{A}} = { m d \ov 4 a}. \eqno(\ref{eq45ef}.b)$$
\input epsf 
 \begin{figure}[as soon as possible]
\begin{center} \leavevmode \hbox{%
\epsfxsize= 8. cm  
\epsfbox{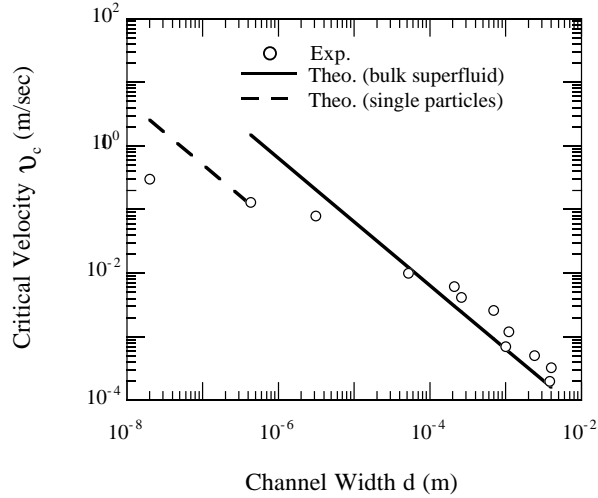}} 
\end{center}
\caption{ 
Critical velocity $v_c(d)$ versus channel width $d$ for He II.  The lines indicate theoretical $v_c(d)$ values predicted from the QCE superfluidity mechanism for a 2D confined superfluid bulk (solid line) and for non-interacting single He atoms of an effective mass (dashed line).  Circles represent experimental data (at about 1.4 \oK) from Ref. \ref{wilks}.
flow, 
\label{fig-vc}  \label{figzII-3} }
\end{figure}

\noindent
Substituting in Eq. (\ref{eq3-7b}) with the expressions (\ref{eq45ef}.b) and (\ref{eq45b})$'$ 
for $\mef$ and $\Delta_{\ph}$ respectively and reorganizing, we obtain the critical velocity
\begin{eqnarray} \label{eq3-9}
 v_{c} =  \sqrt{{48 h c_1 a \ov m}} \quad {1 \ov d} = { 6.42  \times 10^{-7} 
 \over d} \quad \quad {\rm (m/sec)}. \end{eqnarray} 
The parameterization above, where $d$ in meter, is obtained by using $c_1 = 239$ m/sec, $a=3.6 \times 10^{-10}$ m (He II interatomic spacing), and $m=6.64 \times 10^{-27}$ kg ($^4$He atomic mass). The resultant $v_c(d)$ (solid line in Fig. \ref{fig-vc}), containing no adjustable parameters, agrees satisfactorily with the experimental data (circles) for wider channels down to $d \simeq10^{-6}$ m. We accordingly refer to $d<10^{-6} $m as narrower or microscopically narrow channels for which the $v_c(d)$ predicted above deviates from the experimental data, and suggest that the deviation has to do with the fact that $d$ becomes comparable to or less than the correlation length of the superfluid atoms. We can make a rough check, by using the thermal de Broglie wavelength $\L = h /\sqrt{2 \pi m k_B T} \approx 9$ \AA \ at $T=1$ \oK \ as the lower limit of the correlation length, due to only atomic wave overlap[\onlinecite{ess-corr}], $k_B$ being the Boltzmann constant. For $d=10^{-8}$ m, for instance, the $d$ encloses in itself as few number of $\L$'s as $d/ \L \sim 10$. In contrast for a wider channel, of e.g. $d=10^{-4}$ m, $d/\Lambda \sim 1 \times 10^6$.

\noindent
\section{ Superfluidity in narrower channels }\label{Secz5}  \noindent
In a microscopically narrow channel, the superfluid atomic interaction will reduce with $d$. Accordingly the phonon description of the excitation, which is evidently a result of a strong many-atom correlation, 
will become less adequate; rather, the excitation more appropriately involves single atoms.
We assume this is the case with $d \lsim 10^{-6}$ m. Within the small $d$, on the other hand, each helium atom will directly sense the channel boundaries and collisions will primarily occur between a helium atom and the channel wall. To include the two-fold effect above, the atoms may first be approximated as un-correlated but each having an effective (heavier) mass, $m^*$, to account for the small but finite correlation. Second, each single atom is now confined to $d$ in the lateral directions in real space, and confined to a potential well of an infinite depth and a width $d$ in the atomic wavevector $k$ space. The single atom standing wave function for this is $\psi_{\kb}(\rb) = c_{k} e^{i\kb \cdot \rb} +c_{-k} e^{-i\kb \cdot \rb}$. The boundary conditions are $k_{xi}= \i {\widetilde k_x}$, and $k_{zl}= \l {\widetilde k_z}$, $ \i, \l =0, 1, \ldots$; where $ {\widetilde k_x}={\pi \ov d}$ and $ {\widetilde k_z}={\pi \ov L}$. Applying these to the wave function and solving gives the eigen kinetic energy $\epsilon_n = {n^2 h^2 \ov 8 m^* d^2}$, where $n=1,2, \ldots$, for the lateral $k_x$ and $k_y$ states. 
The minimum excitation energy is 
\begin{eqnarray} \label{eq45e} \Delta_{\a} = \epsilon_{2} - \epsilon_{1} =  { 3 h^2 \over  8 m^* d^2 }. \end{eqnarray}
The threshold excitation condition as of Eq. (\ref{eq3-7b}) or (\ref{eq:i:EK}) now applies equally to the atoms, but with the $\Delta_{\ph}$ replaced by the $\Delta_{\a}$. Accordingly, the {\bf QCE superfluidity mechanism} for the single atoms, of a second type here, can be similarly stated as that for the phonons, in terms of the reduction of atomic wavevector states. Similar to the preceding discussions, a critical velocity can be defined when the flow energy $E_s = {1\over 2}N_s m v_s^2 =N_s \e_s$ begins to exceed the (total) threshold excitation energy $N_s\sum_\alpha \Delta_{\a}$ where 
$\alpha=\kxm, \kym,\kzm$ so that $\e_a^c=\sum_\alpha \Delta_{\a}=3 \Delta_{\a}$. Thus 
\begin{eqnarray} \label{eq3-9b} {1\ov 2} m v_c^2 = {9 h^2 \ov 8  m^* d^2}. \end{eqnarray}
 The critical velocity follows to be
\begin{eqnarray}
 v_c (d)= {3h \ov 2 \sqrt{\gamma} md } \simeq  {5.1 \times 10^{-8} \over d}, \end{eqnarray}
where $\gamma =m^*/m$. For the parameterization above, we have made the crude estimation: $\gamma \simeq (\Lambda /a)^3  = (h / a \sqrt{2\pi m k_B T})^3  \simeq 8.6$, with $T=1.4$  \oK. \ Values for other parameters are as specified earlier. The resulting $v_c(d)$ (dashed line in Fig. \ref{fig-vc})
is seen to have an improved agreement with the experimental data in the low $d$ end.  There is however still a noticeable discrepancy; it is evident that first of all, $m^*$ needs to be more realistically determined. Furthermore, as $d$ is reduced to the atomic scale, we expect that the localization of the center of mass of each atom wave in $d$, i.e. that the atomic translation is suppressed to vibrational motion, can result in an even smaller $v_c$.  
\noindent
\section{ Auxiliary comments} \label{Secz6}  \noindent
Supplementary to the preceding discussions, the following points are noted.  
 1). The scarcity of the $K$ modes obtained in this work has no connection to the "Bose-Einstein condensation". As we have noted, the modes dissatisfying Eq. (\ref{eq:apni:7}) are both thermally absent and inhibited from excitation on perturbation; whilst all of those satisfying Eq. (\ref{eq:apni:7}) are nevertheless thermally regularly excited. 2). The well defined discretized excitation state of the superfluid, being an experimental fact and having been shown here to lead to superfluidity, implies a fully coherent atomic motion, which in turn implies relative localization of the atoms in a bulk superfluid. In contrast, the state discretization, which may present at a long wavelength in any fluid, cannot sustain its effect in an essentially diffusive atomic system, as is in the classical fluid or in a bulk quantum fluid which is assumed to be consisting of free atoms. 3). We finally comment that the superfluidity criteria as given by Landau and by Feynman have both incorrectly neglected the primary low energy excitations of phonons. Landau's this ignorance can be traced to the unphysical comparison [\onlinecite{landau}] between velocities of different masses -- the $v_s$ of the correlated superfluid atoms having a mass $(N_s/N_A) m$ where $N_s/N_A > 10^{3}$, and the $c_1$ of a phonon having a zero mass[\onlinecite{mph}];  in the case where a fluid bulk collides with the wall inelastically, such a comparison does not conserve energy . Consistent with his conceptual error, Landau's critical velocity ($\sim$ 100 m/sec) is at least $10^{3}$ times larger than the experimental value and contains no mechanism for $d$ dependence. Feynman in his treatment [\onlinecite{feynman:1953}] (1955) used Landau's critical velocity ($v_L < c_2$, $c_2$ being the phase velocity at $K\approx 1.93$\AA$^{-1}$) as a "valid" starting point, and consequently ignored phonon excitation. Furthermore, Feynman's attribution of turbulence as the source for breaking-up superfluidity suffers the pathology that, for a superfluid assumed to be "non-viscous", turbulence cannot be produced until the turbulence has already been produced. 
\noindent
\section*{\bf Acknowledgements}
\noindent
This work is supported by the Swedish Natural Science Research Council (NFR) and partly by the Wallenbergs Stiftelse (WS); JXZJ specifically acknowledges the postdoctoral fellowships from the NFR and WS. We acknowledge the encouragement from Professor M. Springford, Dr. J. A. Wilson, Dr. P. Meeson, Professor K. Sk\"old, Dr. J. Annett, Dr. O. Ericsson, and Dr. R. Riklund.  
Professor K. Sk\"old has read the final versions of the manuscript and given valuable comments. 
 Since the outline of the (complete) theory of He II in 1998, we have had valuable opportunities of exchanging views with many of the international specialists in liquid ${}^4$He and in superconductivity, and have received encouragements from many of them; we express our deep acknowledgement to them.


\begin{thebibliography}{150}
\bibitem{ess-corr}\label{ess-corr}  
A greater correlation length can be expected due to the presence of a small but finite long range (coulomb) interaction and screening effect of helium atoms upon instantaneous polarization.
\bibitem{mph} Or, in terms of the $m_{\ph}$ of Sec. \ref{Secz1}, $m_{\ph}$ is many orders of magnitudes smaller than the helium atomic mass even for the narrowest channel width in question, $d \sim 10^{-6}$ m.
\end{thebibliography}
\end{document}